\documentclass[sigconf, nonacm]{acmart}
\usepackage[show]{chato-notes}
\usepackage{amsfonts}
\usepackage{acmart-taps}
\usepackage{fontawesome}
\usepackage{xcolor}
\usepackage{framed}
\usepackage{graphicx}

\AtBeginDocument{%
  }

\setcopyright{acmlicensed}
\copyrightyear{2026}
\acmYear{2026}

\acmDOI{XXXXXXX.XXXXXXX}

\acmConference[CIKM '26]{Proceedings of the 35th International ACM Conference on Knowledge and Information Management}{November 7--11, 2026}{Rome, Italy}
\acmBooktitle{Proceedings of the 35th International ACM Conference on Knowledge and Information Management (CIKM '26), November 7--11, 2026, Rome, Italy}
\acmDOI{XXXXXXX.XXXXXXX}
\acmISBN{978-1-4503-XXXX-X/2018/06}

\usepackage{hyperref}
\usepackage{float}
\usepackage{svg}
\graphicspath{ {./images/} }
\begin{document}

\title[Gryphon]{Gryphon: A Unified Architecture for Semantic-ID Generation and Item-Level Scoring in Industrial Recommendations}

\author{Daria Tikhonovich}
\affiliation{%
  \institution{Yandex}
  \city{Moscow}
  \country{Russia}
}
\email{daria.m.tikhonovich@gmail.com}

\author{Oleg Sorokin}
\affiliation{%
  \institution{Yandex}
  \city{Moscow}
  \country{Russia}
}
\email{olegsorokin.ai@gmail.com}

\author{Vladislav Dodonov}
\affiliation{%
  \institution{Yandex}
  \city{Moscow}
  \country{Russia}
}
\email{golafotino@gmail.com}

\author{Mariia Ulianova}
\affiliation{%
  \institution{Yandex}
  \city{Moscow}
  \country{Russia}
}
\email{devchatay@gmail.com}

\author{Ilya Murzin}
\affiliation{%
  \institution{Yandex}
  \city{Moscow}
  \country{Russia}
}
\email{ilyamurzin@yandex-team.ru}

\renewcommand{\shortauthors}{Tikhonovich}

\begin{abstract}

Generative retrieval (GR) has become a scalable approach to candidate generation: each item is assigned a short hierarchical token sequence called a Semantic ID (SID), and the next item's SID is decoded autoregressively. A practical limitation is that the decoder's beam search optimizes the likelihood of token sequences, not the relevance of the underlying items. These objectives diverge when sequence likelihood is poorly calibrated due to beam search error accumulation, and when several items collapse onto a single SID and receive identical scores. We introduce Gryphon, an encoder-decoder generative recommendation architecture that adds a jointly trained item-level scoring component alongside SID generation, reusing the encoder's user representation computed in a single forward pass. Instead of ranking SIDs by accumulated token likelihood, Gryphon resolves each generated SID to its concrete items and re-scores those items directly, which sidesteps miscalibrated sequence scores and separates items that collide on the same identifier. On an industrial music service, with item-level scoring trained under a next-item-prediction objective, Gryphon attains the highest item-level Recall@1000, above the strongest baselines (+3.7\% over vanilla GR and +2.5\% over collision-resolved GR) at comparable parameter count and latency. Gryphon's item-level ranking also surpasses its beam-likelihood ranking of the same candidates (+4.2\% gain), demonstrating the benefit of item-level scoring in GR. Deployed as the sole candidate source in a 7-day A/B test, Gryphon produced no statistically significant change in total listening time (+0.25\%) while replacing a pipeline of more than 15 candidate generators and a separate preranking stage, substantially simplifying the candidate-generation system.

\end{abstract}

\begin{CCSXML}
<ccs2012>
  <concept>
   <concept_id>10002951.10003317.10003347.10003350</concept_id>
   <concept_desc>Information systems~Recommender systems</concept_desc>
  <concept_significance>500</concept_significance>
 </concept>
</ccs2012>
\end{CCSXML}

\ccsdesc[500]{Information systems~Recommender systems}

\keywords{Generative Retrieval, Recommender Systems, Generative Recommendation}




\maketitle

\section{Introduction}

Semantic-ID-based Generative Retrieval (GR) reformulates next-item recommendation as autoregressive generation over a short discrete token sequence—a Semantic ID (SID)—assigned to each item via a hierarchical quantizer such as RQ-VAE~\cite{rajput2023tiger}. Given a user's interaction history encoded by a Transformer encoder, a decoder autoregressively generates the SID of the next item the user is likely to engage with. By casting retrieval as sequence generation over compact semantic identifiers, GR enables end-to-end training and scalable candidate generation in large item corpora, while also leveraging semantic factorization across items~\cite{rajput2023tiger}.

Industrial recommendation, however, is ultimately evaluated at the \emph{item} level, whereas standard GR ranks candidates by accumulated \emph{SID-level} beam likelihood. This creates a structural mismatch that manifests through two failure modes:

\paragraph{Sequence likelihood miscalibration.}
Since beam search conditions on model-generated prefixes rather than ground-truth ones,  early token errors accumulate and leave
relevant items with poorly calibrated SID likelihoods~\cite{ranzato2016sequence,exposurebias2015,promise2026}.

\paragraph{Semantic ID collisions.}
When multiple items share the same SID~\cite{hidvae2025,quasid2026}, beam likelihood assigns them identical scores and cannot express relevance differences among them.

\medskip

Both failure modes reflect the same underlying gap: beam search scores \emph{semantic ids}, while recommendation quality depends on scoring \emph{items}. Recent work has improved SID assignment and highlighted limitations of SID-level decoding and collisions~\cite{hidvae2025,quasid2026,promise2026}, but these approaches still leave open the question of how to rank \emph{resolved items} rather than SID sequences within a unified generative retrieval architecture.

We address this gap with \textbf{Gryphon}\footnote{Gryphon is a project name, not an acronym.}, a unified architecture that augments encoder–decoder GR with a jointly trained \emph{Item-Level Scoring Module} (ILSM). Gryphon performs a single forward pass through a shared encoder; the resulting user states are reused by both an autoregressive decoder for SID generation and the ILSM, which scores the concrete items resolved from the generated SIDs. By re-ordering items with item-level scores rather than beam likelihood, Gryphon reduces dependence on potentially miscalibrated sequence-level scores and distinguishes among items sharing the same SID.

\paragraph{Contributions.}
\begin{itemize}
\item We identify a structural mismatch in SID-based generative retrieval — beam search ranks SID sequences, not items — and characterize two resulting failure modes: sequence likelihood miscalibration and Semantic ID collisions.

\item We propose Gryphon, a shared-encoder architecture jointly training SID generation and item-level scoring under roughly matched parameter count and inference budget, decoupling final item selection from beam likelihood.

\item We validate Gryphon on a large-scale industrial music platform. Offline, it achieves the highest item-level Recall@1000 of all evaluated baselines under roughly matched parameter count and inference budget, and we show that item-level re-scoring exceeds the SID-level beam ceiling — isolating beam-likelihood miscalibration, not candidate recall, as the limiting factor. In a 7-day online A/B test, as
the sole candidate source it produced no statistically significant change in total listening time while passing 66.7\% fewer candidates to the final ranker and removing the preranking stage entirely.

% \item We validate Gryphon in a large-scale industrial music setting. Offline, it improves Recall@1000 by +3.7\% and Recall@10 by +11.2\% relative over vanilla GR. In a 7-day online A/B test, Gryphon as the sole candidate source matches the listening time of a production stack with 15+ candidate generators and a preranker, while passing 66.7\% fewer candidates to the final ranker.
\end{itemize}

\section{Background and Motivation}
\subsection{SID-Based Generative Retrieval}
Semantic-ID-based generative retrieval represents each item $x\in\mathcal{X}$ by a short sequence of discrete tokens and trains an encoder--decoder model to generate the identifier of the next item from the user history~\cite{rajput2023tiger,hidvae2025,quasid2026,ju2026semantic}. Formally, a quantizer maps each item to a $d$-token Semantic Identifier (SID),
$$
\Phi:\mathcal{X}\to\prod_{b=1}^{L}\{1,\dots,M_b\}, \qquad
\Phi(x)=(s_1,\dots,s_L).
$$
Given user history $u$, GR autoregressively predicts candidate SIDs. For a candidate SID $\sigma=(s_1,\dots,s_L)$, beam search ranks candidates by the accumulated code-level likelihood~\cite{wu2016gnmt,rajput2023tiger}
$$
\ell_\theta(\sigma\mid u)=\sum_{b=1}^{L}\log p_\theta(s_b\mid u,s_{<b}).
$$
This score is natural for decoding identifiers, but it is defined over token sequences, whereas recommendation quality is evaluated over concrete items~\cite{covington2016youtube,liu2022neuralreranking}.

\subsection{Semantic-ID Collisions}
\label{sec:collisions}
Semantic-ID collisions make the scoring gap strict when multiple items share the same generated identifier~\cite{hidvae2025,quasid2026,ju2026semantic}. Let $\mathcal{C}_\sigma=\Phi^{-1}(\sigma)$ be the collision group for SID $\sigma$, i.e., the set of concrete items sharing $\sigma$. For any two items $x_i,x_j\in\mathcal{C}_\sigma$,
$$
\ell_\theta(\Phi(x_i)\mid u)=\ell_\theta(\Phi(x_j)\mid u),
$$
because both items map to the same SID. Beam likelihood therefore assigns identical scores to all items in the collision group and cannot express relevance differences among them through the SID score alone. Collisions expose a structural limitation of pure SID-level scoring: the generator scores identifiers, while the recommender must rank concrete items. Previous work~\cite{rajput2023tiger} removes collisions by appending an extra token to the ordered semantic codes to make them unique. We argue this is better suited to static offline settings: a unique terminal token per colliding item ties the resolution vocabulary to the catalog, so under dynamic catalogs with a continual influx of new items the resolving layer must grow and be re-fit indefinitely---a serious obstacle to deployment. 

\subsection{From SID Likelihood to Item-Level Relevance}
Autoregressive generation is trained with teacher forcing, yet inference conditions on the model's own prefixes~\cite{exposurebias2015,ranzato2016sequence}; in hierarchical SID decoding, an early prefix error can move the beam into a different subtree, leaving relevant items on lower-scored SID paths~\cite{zhao2023slic,promise2026}. Beam search is thus a strong candidate-generation mechanism, but beam likelihood alone is not a complete item-level relevance signal. This motivates Gryphon's jointly-trained Item-Level Scoring Module, introduced in Section~\ref{sec:method}.

%Gryphon therefore treats generated SIDs as candidates to be resolved into items, then scores those items with a target-item-aware module.

% This design addresses the mismatch between SID-level generation and item-level
% recommendation~\cite{covington2016youtube,liu2022neuralreranking}. In vanilla GR, beam search scores SID sequences using
% autoregressive likelihoods, although the final recommendation target is an
% item. Gryphon instead treats beam search as a candidate-generation mechanism:
% generated SIDs are expanded into their corresponding item collision groups, and
% items in the resulting candidate pool are scored by an item-level module. 
% At inference time, beam search generates a set of candidate SIDs~\cite{wu2016gnmt,rajput2023tiger}, which defines the candidate pool that is subsequently scored at the item level.

% The ILSM is used as a retrieval-stage scoring function over candidate items
% rather than as a globally calibrated probability estimator. We therefore
% instantiate it with a next-item prediction objective implemented as a sampled
% contrastive loss, which encourages the observed next item to receive a higher
% score than sampled negatives.

\section{Methodology}\label{sec:method}

\begin{figure}[t]
    \centering
    \includegraphics[width=0.5\textwidth]{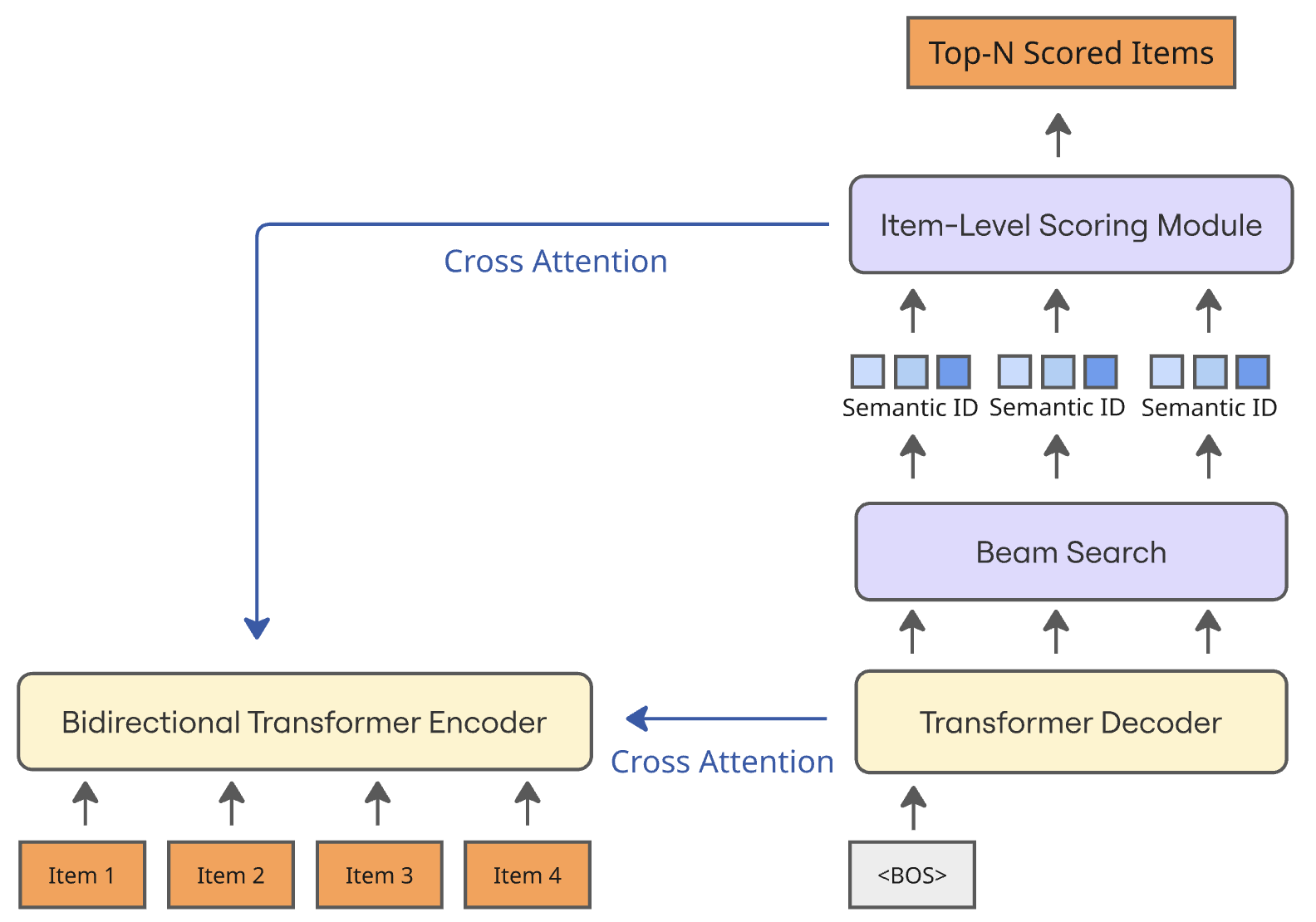}
    \caption{Gryphon inference: beam search generates candidate SIDs, which are resolved to all matching items and re-scored with ILSM, while beam likelihood is discarded from ranking}
    \label{fig:Gryphon}
\end{figure}

Gryphon augments a vanilla encoder--decoder generative retriever with an
Item-Level Scoring Module (ILSM). For each user request, the user behavior
sequence is encoded once into encoder states \(E_u\). These
states are shared by both components: the autoregressive decoder uses \(E_u\)
to generate SID candidates, while the ILSM reuses the same
\(E_u\) to score items associated with the generated SIDs.

As a result, final item scores are not tied to potentially miscalibrated sequence likelihoods, and Gryphon can distinguish among items within the same SID collision group.

\subsection{Generative Retrieval}
The decoder autoregressively predicts the SID of the next item given encoded user history. For an observed next
item \(i^+\) with SID \(\Phi(i^+)=(s_1^+,\ldots,s_L^+)\), the generation loss is
\[
\mathcal{L}_{\mathrm{gen}}
=
-\sum_{t=1}^{L}
\log p_\theta(s_t^+ \mid s_{<t}^+, E_u).
\]

\subsection{
%Item-Level Cross-Attention Ranker
Item-level Scoring Module %Cross-Attention Scorer
}

To alleviate the inconsistency between SID generation and item-level recommendation, we introduce an item-level scorer for generative recommendation. Each candidate item \(i\) is represented by an item-query
embedding produced by an item tower~\cite{covington2016youtube,singh2024better, zhao2023embedding}:
\[
e_i = \mathcal{T}_{\mathrm{item}}(\Phi(i), h_i),
\]
where \(h_i\) denotes native item-level features, such as item-ID hashes,
metadata and content features. The scorer then conditions this item query on
the shared user encoder states \(E_u\):
\[
r_\phi(u,i) = f_\phi(E_u, e_i).
\]
In practice, \(f_\phi\) is a lightweight item-to-user cross-attention
block followed by an MLP head that outputs a scalar relevance score.

The ILSM directly addresses SID collisions~\cite{hidvae2025,quasid2026,ju2026semantic}, since items with the same SID can
still receive different relevance estimates through their item-level features. The ILSM also mitigates, but does not eliminate, error accumulation in
autoregressive decoding. Beam search still determines which SIDs enter the
candidate pool. However, once a SID has been generated, the final item selection
no longer depends on the product of SID-token likelihoods. This reduces the
effect of poorly calibrated sequence likelihoods among items reachable through
the generated SID set.

\subsection{Training Objective}

The ILSM is not architecturally tied to a particular supervision signal.
It can be trained with a variety of item-level objectives, including multi-objective engagement prediction, ranker distillation, or long-term value optimization.
In this work, we instantiate the ILSM with a next-item prediction objective optimized using sampled softmax, and leave richer item-level objectives to future work.
Thus, the ILSM is trained to maximize the probability of item \( p_{\phi}(i_{t+1} = i \mid E_u) \propto \exp\left( \frac{r_{\phi}(u, i)}{\tau} \right)
 \ \), conditioning on the user history.

Given a user \(u\), the observed next item \(i^+\), and a set of in-batch sampled
negative items \(\mathcal{B}^-\), we define 
\(
\mathcal{L}_{\mathrm{NIP}}
\):
\begin{align*}
% &\mathcal{L}_{\mathrm{NIP}} 
\\
&-\log
\frac{
\exp\left(\frac{r_\phi(u,i^+)}{\tau} - \log Q_{i^+}\right)
}{
\exp\left(\frac{r_\phi(u,i^+)}{\tau} - \log Q_{i^+}\right)
+
\sum_{i^- \in \mathcal{B}^-}
\exp\left(\frac{r_\phi(u,i^-)}{\tau} - \log Q_{i^-}\right)
},
\end{align*}
where \(\tau\) is a temperature parameter and \(Q_i\) is the sampling
probability of item \(i\). The sampling correction term reduces popularity-induced bias \cite{yi2019sampling}.

The full Gryphon objective jointly trains SID generation and item-level scoring:
\(
\mathcal{L}
=
\mathcal{L}_{\mathrm{gen}}
+
\lambda \mathcal{L}_{\mathrm{NIP}},
\)
where \(\lambda\) controls the contribution of the item-level loss. Both losses
share the same encoder states \(E_u\), encouraging the user representation to
support both SID-level candidate generation and item-level relevance estimation.

\subsection{Inference}

At serving time, Gryphon performs a single encoder forward pass to compute
\(E_u\). The decoder then uses beam search to generate a top-\(K\) set of SID
candidates, denoted by \(\mathcal{B}_u\). Beam likelihoods are used only to
determine membership in \(\mathcal{B}_u\), they are not used as final item
scores. 
This choice intentionally decouples item selection from potentially miscalibrated SID-token products, while still using the decoder to propose a compact SID set.

Each generated SID is expanded to its collision group, and Gryphon then scores
the resulting items with the ILSM:
% \[
% \mathrm{TopN}(u)
% =
% \operatorname{TopN}_{i \in \bigcup_{\sigma \in \mathcal{B}_u} \mathcal{C}_\sigma}
% r_\phi(u,i).
% \]
\[
\mathrm{TopN}(u)
=
\operatorname{TopN}_{i \in \mathcal{I}_u} r_\phi(u,i),
\qquad
\mathcal{I}_u
=
\left\{
i \;\middle|\;
\exists \sigma \in \mathcal{B}_u \text{ s.t. } i \in \mathcal{C}_\sigma
\right\}.
\]
The resulting Gryphon candidate pool can be passed to a downstream ranker for final action prediction.

\section{Experiments}
% \subsection{Setup}

Our experiments aim to address the following research questions:

\textbf{RQ1}. How does Gryphon compare to Transformer-based production baselines and SID-based GR baselines on item-level next-item prediction?

\textbf{RQ2}. Can Gryphon replace the deployed production candidate-generation stack without degrading online engagement?

\subsection{Offline Evaluation }

\subsubsection{Dataset}

We collect interactions from one week of real user logs from a large-scale music recommendation platform with tens of millions of active users and items.

\subsubsection{Baselines}

We compare Gryphon against:

\textbf{ARGUS.} ARGUS~\cite{khrylchenko2025scaling} is a deployed autoregressive two-tower Transformer candidate generator trained with sampled softmax for sequential next-item prediction~\cite{hidasi2016gru4rec,kang2018sasrec}. Being among the largest recommendation models currently in service, ARGUS serves as a strong production baseline.

\textbf{Vanilla Generative Retrieval.} This baseline represents target items as SID sequences and generates them autoregressively. Since multiple items may share the same generated sequence, we resolve collisions by expanding generated identifiers in beam order and accumulating associated items until the candidate budget is reached. Since collisions may pessimize item-level recall, we also implement a Resolved variant for vanilla GR where we append an extra token at the end of SID sequence, following \cite{rajput2023tiger}. We present the Resolved GR baseline for offline comparison only, since the resolving layer can grow unbounded under production catalog changes and we find this model impractical for our deployment setting.

\subsubsection{Implementation Details}

For SID-based models we quantize all items into SIDs using residual $K$-Means. We independently tune the number of codebooks from 1 to 4, codebook sizes from 1,024 to 32,000, and the encoder-decoder layers ratio for each model. For ARGUS we use $10$ transformer encoder blocks. For Vanilla GR, we use $7$ encoder blocks, $3$ decoder blocks, and 3 codebooks with a size of 32,000. Compared to vanilla GR, for Gryphon, we use the same codebook setup and replace one decoder block with a single-layer ILSM, resulting in less than 1\% difference in parameter count and inference time. In the ILSM we use only the item-id feature, so that Gryphon has no item-feature advantage over the GR baselines. We set $\lambda$ to 1. For vanilla GR Resolved, we use $9$ encoder blocks, $1$ decoder block, and 2 codebooks with a size of 1024, as we found that larger codebooks greatly degrade the performance of this baseline. We use a beam size of $K=2048$, user sequence length of $512$, and hidden size of $1024$ for all models. We use in-batch sampling with cross-device gathering as a source of negatives for LogQ-corrected Sampled Softmax for ARGUS and ILSM. All models are trained using Adam with linear warmup followed by linear learning-rate decay and use identical features for encoder inputs.

\subsubsection{Metrics}

Recall@$k$ measures the fraction of ground-truth items associated with a user impression that appear among the top-$k$ returned items. Because the retrieved candidates are subsequently processed by a production ranker, Recall@1000 is the primary offline metric for measuring candidate-generator quality.
Due to the high computational cost of repeated experiments, we do not perform formal statistical significance testing across multiple random seeds. As a practical estimate of stochastic variability, we conducted a limited set of preliminary experiments with different random initializations and observed that Recall@10 and Recall@1000 varied by approximately ±0.003. We therefore interpret differences of this magnitude or smaller with caution.

\subsubsection{Results}

Experimental results are reported in Table~\ref{tab:offline-item-results}.
Gryphon achieves the highest item-level Recall@$k$ at both evaluated cutoffs. Gryphon  substantially outperforms the production ARGUS baseline.
Compared with Vanilla GR, Gryphon improves Recall@1000 from 0.8245 to 0.8552. Compared with Vanilla GR Resolved,
Gryphon improves Recall@1000 from
0.8343 to 0.8552. These gains exceed our empirical initialization-variance
estimate of approximately $\pm 0.003$, although we do not perform formal
multi-seed significance testing.

To isolate the source of Gryphon's gain, we score an identical
($K{=}2048$) beam-generated SID pool three ways
(Table~\ref{tab:ilsm-ablation}). Under beam-likelihood ranking,
item-level Recall@1000 ($0.8209$) falls below the model's SID-level
Recall@1000 ($0.8404$), quantifying the cost of collision resolution:
beam likelihood cannot distinguish items sharing a SID. ILSM
re-scoring of the same pool raises item-level Recall@1000 to
$0.8552$, \emph{above} the SID-level beam ceiling. Since item-level
recall can exceed this ceiling only by promoting items from SIDs the
beam ranked beyond its top $1000$, the crossing shows that beam
likelihood miscalibrates \emph{reachable} candidates and that
item-level scoring corrects it.

% Gryphon gain comes from item-level scoring rather
% than the generative backbone---beam likelihood, not the candidate
% pool, is the limiting factor.

% To isolate the ILSM's contribution, we report \textbf{Gryphon w/o ILSM}, the same jointly trained model ranked by beam likelihood rather than ILSM scores. This variant reaches only 0.8209 Recall@1000, slightly below Vanilla GR (0.8245), since one decoder block is replaced by the ILSM. ILSM re-scoring of the same candidate set raises Recall@1000 to 0.8552, confirming that the gain comes from item-level scoring rather than the generative backbone, and that beam likelihood miscalibration — not the candidate pool — is the limiting factor.

\begin{table}[t]
\centering
\caption{Offline item-level next-item prediction results.}
\label{tab:offline-item-results}
\begin{tabular}{lccc}
\hline
\textbf{Method} & \textbf{SIDs} & \textbf{Recall@10} & \textbf{Recall@1000} \\
\hline
ARGUS & - & 0.0996  & 0.6582 \\
Vanilla GR & 3x32000  & 0.1961  & 0.8245 \\
Vanilla GR Resolved & 2x1024   & 0.2077  & 0.8343 \\
% Full Softmax          & 0.1530  & 0.8230 \\
Gryphon (ours)  &  3x32000     & \textbf{0.2178}  & \textbf{0.8552} \\
% \midrule
% Gryphon w-o ILSM  &  3x32000     & 0.1849  & 0.8209 \\
\hline
\end{tabular}
    % {\footnotesize $\pm 0.003$ is our empirical estimate of initialization variance}
\end{table}

\begin{table}[t]
  \centering
  \caption{ILSM ablation on a shared candidate pool ($K{=}2048$).}
  \label{tab:ilsm-ablation}
  \begin{tabular}{llc}
    \toprule
    Variant & Scoring & Recall@1000 \\
    \midrule
    Gryphon w/o ILSM & SID-level (beam scores)              & 0.8404 \\
    Gryphon w/o ILSM & item-level (beam scores) & 0.8209 \\
    Gryphon   & item-level (ILSM)       & \textbf{0.8552} \\
    \bottomrule
  \end{tabular}
\end{table}

%Thus, within the sensitivity of this experiment, replacing the full production candidate-generation stack with Gryphon did not produce a measurable degradation in the primary engagement metric. At the same time, Gryphon reduced the number of candidates passed to the final ranker by 66.7\% and eliminated the separate preranking stage.

\subsection{Online A/B Test Results}

We evaluated Gryphon in a 7-day online A/B experiment against the production candidate generation stack in a music recommendation application. Users were randomly assigned to control and treatment buckets, with 4\% of eligible users allocated to the treatment. The control arm used the existing production candidate-generation pipeline, consisting of more than 15 heterogeneous candidate generators. These generators produced 10{,}000 candidates per request, which were filtered by a production preranker before 3{,}000 candidates were passed to the final ranker. In the treatment arm, this candidate-generation and preranking stack was replaced by Gryphon. Gryphon served as the sole candidate source and directly supplied 1{,}000 candidates to the same final ranker.

Table~\ref{tab:ab} summarizes the configuration and results. The primary engagement metric in the service is total listening time (TLT). Gryphon changed TLT by +0.25\% relative to the production control, and the difference was not statistically significant within our experimental design. On secondary quality metrics, Gryphon showed statistically significant favorable movement: a higher active-user ratio (+0.43\%), and fewer unfinished tracks (-1.3\%).  We interpret these results as evidence that Gryphon can serve as the sole candidate source without a statistically significant change in primary engagement while substantially simplifying the candidate-generation and preranking pipeline.

\begin{table}[tb]
    \centering
    \caption{Online A/B test results.}
    \label{tab:ab}
    \begin{tabular}{lcc}
        \toprule
          &
        \textbf{Production stack} &
        \textbf{Gryphon} \\
        \midrule
        Candidate generators & 15+ & 1 \\
        Initial candidates & 10{,}000 & 1{,}000 \\
        Preranking stage & Yes & No \\
        Candidates passed to ranker & 3{,}000 & 1{,}000 \\
        \midrule
      Total listening time & -- & $+0.25\%$ \\
      Active users ratio   & -- & $+0.43\%^{*}$ \\
      Unfinished tracks    & -- & $-1.3\%^{*}$ \\
      % Dislikes             & -- & $-5.7\%^{*}$ \\
        \bottomrule
    \end{tabular}
{\footnotesize\normalfont
Changes relative to control. $^{*}$ Significant at $p < 0.001$.
}
\end{table}

\section{Conclusion and Future Work}
We presented Gryphon, a unified architecture for combining Generative Retrieval with item-level scoring in generative recommendations. Gryphon treats beam search as SID-level candidate generation, resolves generated SIDs into concrete items, and applies an Item-Level Scoring Module to re-rank the resolved candidates. By separating SID-sequence likelihood from item-level relevance across GR and ILSM, Gryphon reduces dependence on potentially miscalibrated sequence-level
scores, distinguishes between items within SID collisions and utilizes item-level features for final scoring.

Offline experiments show that Gryphon with a next-item prediction training objective improves retrieval quality over a production-compatible vanilla GR baseline under roughly matched parameter count and inference budget. In a 7-day online A/B test, replacing a full production candidate-generation and preranking stack with Gryphon as the sole candidate source produced no statistically significant change in total listening time, while reducing the number of candidates passed to the final ranker from 3{,}000 to 1{,}000.

Although Gryphon substantially simplifies the candidate-generation stack, it does not eliminate the need for a production ranker under the tested next-item prediction objective. Future work includes extending the Item-Level Scoring Module to richer training objectives, such as multi-objective engagement modeling, distillation from production rankers, and long-term user-value optimization.

\section*{GenAI Usage Disclosure}
Generative AI tools were used to support manuscript revision, including English-language editing, clarity improvements, and high-level feedback on the presentation before submission. The authors manually reviewed and edited all AI-assisted suggestions and take full responsibility for the final manuscript.

\bibliographystyle{ACM-Reference-Format}
\bibliography{references}

\end{document}